\documentclass[pra,preprint,showpacs,amsmath,amssymb,superscriptaddress]{revtex4}
\usepackage{graphicx}
\usepackage{natbib}
\usepackage{latexsym}

\begin{document}

\title{Modifying molecular scattering from rough solid surfaces using ultrashort laser pulses}

\author{Yuri Khodorkovsky}
\affiliation{Department of Chemical Physics, The Weizmann Institute of Science, Rehovot 76100, Israel}

\author{J. R. Manson}
\affiliation{Department of Physics and Astronomy, Clemson University, Clemson, SC 29634, USA}

\author{Ilya Sh.\ Averbukh}
\affiliation{Department of Chemical Physics, The Weizmann Institute of Science, Rehovot 76100, Israel}



\begin{abstract}
We consider solid surface scattering of molecules that were subject to strong non-resonant ultrashort laser pulses just before hitting the surface. The pulses modify the rotational states of the molecules, causing their field free alignment, or a rotation with a preferred sense. We show that field-free laser induced molecular alignment leads to correlations between the scattering angle and the sense of rotation of the scattered molecules. Moreover, by controlling the sense of laser induced unidirectional molecular rotation, one may affect the scattering angle of the molecules. This provides a new means for separation of mixtures of molecules (such as isotopes and nuclear-spin isomers) by laser controlled surface scattering.
\end{abstract}

\maketitle


\section{Introduction}
\label{intro}

Laser control of molecular rotation, alignment and orientation has received significant attention in recent years (see e.g.\ reviews \cite{Stapelfeldt03}). Interest in the field has increased, mainly due to the improved capabilities to manipulate the characteristics of the laser pulses (such as time duration and temporal shape), which in turn lead to potential applications offered by controlling the angular distribution of molecules. Since the typical rotational time scale of small molecules is 'long' (${\sim}10 \,\mathrm{ps}$) compared to the typical short pulse duration (${\sim}50 \,\mathrm{fs}$), effective rotational control and manipulation are in reach.
During the last decade, coherent rotational dynamics of pulse-excited molecules was studied \cite{Ortigoso}, and multiple pulse sequences giving rise to enhanced alignment were suggested \cite{Averbukh}, and realized experimentally \cite{Bisgaard1}. Further manipulations, such as optical molecular centrifuge \cite{Karczmarek} and alignment-dependent strong field ionization of molecules \cite{Litvinyuk}, were demonstrated. Selective rotational excitation in bimolecular mixtures was suggested and demonstrated in the mixtures of molecular isotopes \cite{Fleischer06} and molecular nuclear-spin isomers \cite{Renard04}.
Aligned molecules were also shown to be useful for other applications, including: generation and modification of ultrashort light pulses \cite{Bartels}, control of high harmonic generation \cite{Velotta}, amplification of terahertz pulses \cite{Milchberg}, elucidation of molecular structure \cite{Itatani}, rotational spectroscopy \cite{Riehn}, manipulation of chemical reactions \cite{Larsen}, and others.
Finally, it was demonstrated that these new methods for manipulation of molecular rotation can also be used to modify the motion of molecules in inhomogeneous fields, such as focused laser beams \cite{Purcell09,frisbee}, or static electric \cite{electric} and magnetic \cite{magnetic} fields.

Modification of the molecule-surface scattering and molecule-surface reactions by external fields of different nature is a long-standing research problem. In particular, the effect of molecular orientation by a static electric hexapole field on the scattering process was investigated in detail \cite{Kuipers91}. Laser control of the gas-surface scattering process was achieved using multiphoton ionization of the impinging molecules by long laser pulses of variable polarization \cite{Jacobs}, and the possibility of controlling  molecular {\it adsorption} on solid surfaces using ultrashort laser pulses was discussed \cite{Seideman_surf}. 

In a recent paper \cite{Khodorkovsky11}, we investigated two possible schemes for modifying molecular scattering from {\it flat} surfaces. In the first scheme, we considered excitation of unidirectional molecular rotation via the combined action of a short laser pulse on molecules in a molecular beam, followed by a surface collision of these molecules. In the second scheme, we suggested exciting a molecular beam, consisting of two molecular species, by two properly delayed laser pulses, in order to provide selective rotational excitation of one of the species. We have shown that these two species are then scattered differently from a solid surface, in a way that the scattered beam is enriched in one of the species, for specific scattering angles.

The current paper extends the model of molecule-surface scattering that we used in \cite{Khodorkovsky11}. We treat here surfaces that are not flat, but {\it corrugated} on different length scales. One limit includes surfaces whose corrugation period is larger than the size of the molecule. The model we developed according to this limit is based on the model in \cite{Tully90}. This model is referred to throughout the paper as the {\it corrugated surface} model. This limit corresponds to highly corrugated surfaces, such as $\text{Pt}(211)$ \cite{Tully90}, or $\text{LiF}(001)$ \cite{Kondo05}. Another limit includes surfaces with corrugation period smaller than the size of the molecule, or relatively flat surfaces. Here the model is based on the work of \cite{Sitz88}. This model is referred to below as the {\it surface with friction} model. This limit corresponds to relatively flat surfaces, such as $\text{Ag}(111)$, where mechanisms other than surface corrugation exist for producing forces tangent to the surface, such as: tangentially directed phonons, creation of electron-hole pairs, etc. \cite{Sitz88}.

In this paper, we continue to investigate the ability to modify and to control the molecule-surface scattering processes by using ultrashort laser pulses. These laser pulses (mainly femtosecond laser pulses) align the molecules or set them in some specific rotational motion, before the molecules hit the surface.
The paper is organized as follows: In Sec.\ \ref{model1} we present the model of molecular scattering from a corrugated surface. The classical model of interaction between an ultrashort laser pulse and a linear rigid molecule is briefly summarized in Sec.\ \ref{laser}. Then, in Sec.\ \ref{results1}, it is shown that a laser pulse aligning the molecules before they scatter from a corrugated surface, leads to a correlation between the direction of scattering of the molecules and their sense of rotation. This correlation also depends on the direction of polarization of the laser. The two-dimensional and the three-dimensional models of molecular scattering from a surface with friction is next presented in Sec.\ \ref{model2}. Accordingly, in Sec.\ \ref{results2}, the scattering of molecules from a surface with friction, set in rotation in a specific sense by laser pulses, is investigated. It is shown, that a correlation exists between the sense of rotation of the molecules impinging on the surface, and their scattering angles. Finally, in Sec.\ \ref{conclusions} we conclude and summarize.

\section{A model of molecular scattering from a corrugated surface}
\label{model1}

Here we develop a model and analyze the scattering of a molecule from a surface that is corrugated at a length scale larger than the size of the molecule. This model can qualitatively describe the scattering process from surfaces, such as Pt(211) \cite{Tully90}. 

The model developed here is an extension of our model \cite{Khodorkovsky11} from flat to corrugated surfaces. Our treatment of corrugations is based on the model of Tully \cite{Tully90}. The diatomic molecule is treated as a rigid dumbbell \cite{Doll73}. This dumbbell collides with a corrugated, but {\it frictionless}, surface, which is composed of hard cubes on identical springs, see Fig.\ \ref{corrug_pic}. Each cube represents one (or several) of the surface atoms \cite{Logan66,Tully90,Kondo05}. We assume that the corrugation is one dimensional, and that its period is larger than the molecular bond length, such that the molecule collides with a single cube that is oriented in a definite angle. We also assume that the cube oscillates with an amplitude determined by the surface temperature. This hard cube model provides a simple way of adding surface phonons to the process. For the sake of simplicity, we also assume that the cube is much heavier than the molecule, so that its velocity does not change as a result of the collision. By moving to the reference frame attached to the cube, one reduces the problem to the molecular collision with a motionless hard wall. In the moving coordinate system, the molecular total energy (translational+rotational) is conserved, but it can be redistributed between these two parts as a result of the collision.

What about the translational linear momentum of the center of mass of the molecule in the moving coordinate system? The momentum component that is {\it locally perpendicular} to the surface is not conserved, because the vibrating cube exerts forces on the molecule in this direction during the collision. On the other hand, there are no forces applied in the direction that is {\it locally parallel} to the frictionless surface, and therefore the linear momentum locally parallel to the surface is conserved. It follows, that because of the corrugation, the momentum that is {\it globally parallel} to the surface is not conserved, contrary to the case of a flat frictionless surface.

Using energy and angular momentum conservation laws (angular momentum is conserved only in the coordinate frame centered at the position of the colliding atom), we find analytic expressions for the translational and the rotational velocities of the dumbbell molecule after the collision. These velocities depend on the local velocities before the collision, and on the local angle between the dumbbell and the face of the cube at the moment of collision. Finally, we transform the velocities back to the laboratory coordinate frame.

It is important to emphasize that our model treats the molecule as a dumbbell, although it is also possible to treat it as an ellipsoid \cite{Kondo05}. In our model, differently from the model used in \cite{Kondo05}, we include three-dimensional rotation and multiple collisions of the molecule with the surface.

In the next subsections, we first define the coordinates and the surface corrugation in subsection \ref{coordinates1}.
Then, we consider the effects of the surface-cube vibration on the collision in subsection \ref{vibration}.
In subsection \ref{collision1}, we provide the results for the molecular velocities after the collision as a function of the velocities before the collision and of the molecular orientation.

\subsection{The coordinates and the one-dimensional surface corrugation}
\label{coordinates1}

\begin{figure}
\centering
\includegraphics[width=1\textwidth]{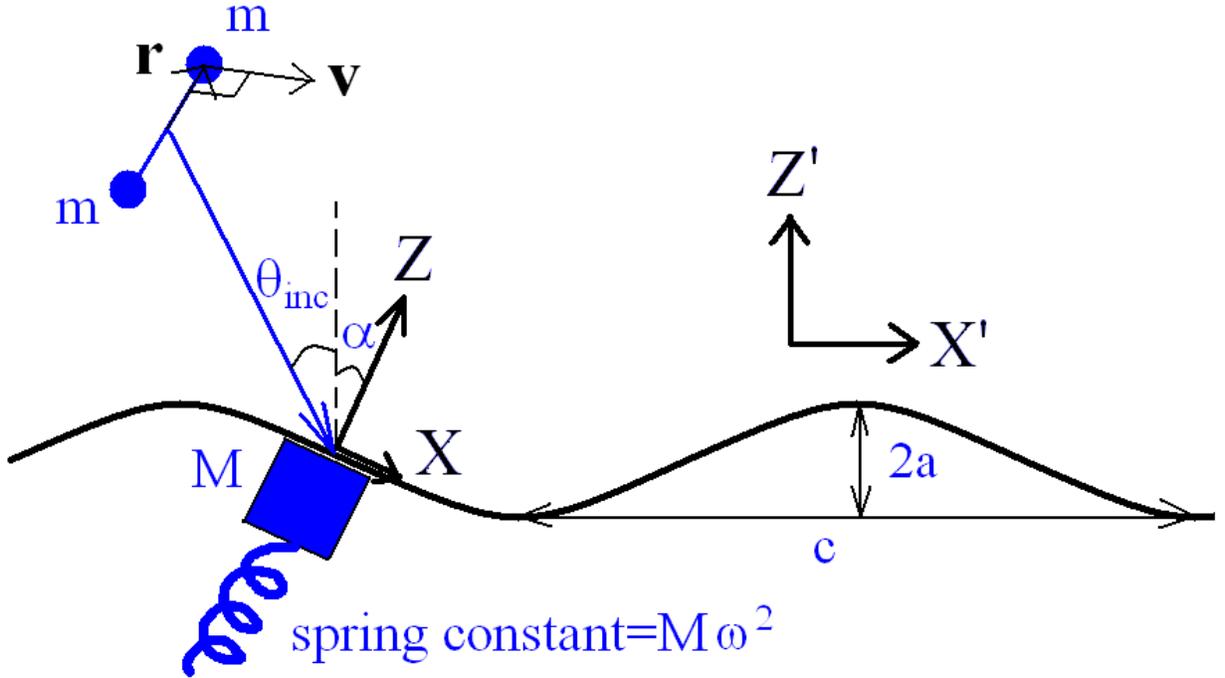}
\caption{Schematic drawing of the dumbbell -- hardcube corrugated surface model. The surface is a two-dimensional sinusoidal sheet composed of small hard cubes of mass $M$. The period of the sine corrugation is $c$ and its amplitude is $a$. The cubes are held by springs that have a spring constant of $M\omega^2$. They oscillate in a direction perpendicular to the local surface tangent, denoted by the $Z$-axis. The inclination of the $Z$ axis from the global surface normal (shown by the dashed line) is denoted by $\alpha$. The molecule is a massless stick with two balls of mass $m$ on its ends. The molecular orientation of one of the atoms is given by the vector $\mathbf{r}$, and the linear velocity of this atom is $\mathbf{v}$. The incident molecule approaches the surface at an angle of incidence of $\theta_{\text{inc}}$, which is always negative. The global coordinate system $X'Z'$ is shown on the right.}
\label{corrug_pic}
\end{figure}

We treat a homonuclear diatomic molecule as a massless stick of length $r_e$ (equal to the bond length), with two balls, each of mass $m$, attached to its ends.
To describe the motion of the molecular center of mass, we define the $Z'$- and $X'$-axes as {\it globally} perpendicular and parallel to the surface, respectively. We also define the $Z$- and $X$-axes as {\it locally} perpendicular and parallel to the surface, respectively, see Fig.\ \ref{corrug_pic}. The {\it local} molecular orientation is given by the vector $\mathbf{r}=0.5r_e (\sin{\theta}\cos{\varphi}, \sin{\theta}\sin{\varphi}, \cos{\theta})$. Notice that this molecular orientation depends on the orientation of the cube on which the molecule falls. Here the polar angle between the molecular axis and the $Z$-axis is $\theta$, ranging from $0$ to $\pi$, and $\varphi$ is the azimuthal angle, with values from $0$ to $2\pi$. The {\it global} translational velocity components of the center of mass are denoted by $V_Z'$ and $V_X'$. The first component is defined as positive if the molecule moves upwards, and the second one is positive if the molecule moves to the right. The {\it local} translational velocity components (in the unprimed coordinate system) are denoted by $V_Z$ and $V_X$. The linear rotational velocity is denoted by $\mathbf{v}$, and is equal to $d\mathbf{r}/dt$. 

The incident and the scattering angles are measured with respect to the $Z'$-axis, and are positive in the clockwise direction and negative in the counter-clockwise direction. The angle of incidence of the molecular beam is denoted by $\theta_\text{inc}$, and is always chosen as non-positive. The angle between the {\it local normal} to the surface $Z$ and the $Z'$-axis is denoted by $\alpha$.
The velocities in the primed coordinates can be expressed as a function of the velocities in the unprimed coordinates and the angle $\alpha$:
\begin{equation}
\label{transformation}
\left( 
\begin{matrix}
V_X \\
V_Z 
\end{matrix}
\right)
=
\left(
\begin{matrix}
\cos{\alpha} & -\sin{\alpha} \\
\sin{\alpha} & \cos{\alpha} 
\end{matrix}
\right)
\left(
\begin{matrix}
V'_X \\
V'_Z  
\end{matrix}
\right)~. 
\end{equation}

Following the work of Tully \cite{Tully90}, we choose the simplest sinusoidal one-dimensional corrugation of the form:
\begin{equation}
\label{corrugation}
Z'(X')=a\cos{\left(\frac{2\pi X'}{c}\right)}~, 
\end{equation}
where $a$ is the amplitude of the corrugation, and $c$ is its period. Scattering out of the $X'Z'$-plane is not possible in the framework of this model, due to the absence of corrugation along the $Y'$-direction, and to the absence of a local friction. This treatment corresponds to the measurement of in-plane scattering \cite{Pollak11}. The model used in Sec.\ \ref{model2} incorporates out-of-plane scattering due to frictional forces parallel to the surface.

The tangent of the angle of the local surface normal $\alpha$ is given by:
\begin{equation}
\label{tan_alpha}
\tan{\alpha}=-\frac{dZ'}{dX'}=\frac{2\pi a}{c}\sin{\left(\frac{2\pi X'}{c}\right)}~. 
\end{equation}
The maximal local normal angle is $\alpha_{\text{max}}=\arctan{\left(2\pi a/c\right)}$, which is referred to as corrugation strength. A molecule incident on the surface at an angle $\theta_\text{inc}$, hitting a cube inclined at $\alpha$, has an effective incident angle of $\theta_\text{inc}-\alpha$. We will always assume that no ``shadowing'' occurs, which means that $\theta_\text{inc}-\alpha_{\text{max}}>-\pi/2$. We also assume that after a collision (possibly multiple) with a single cube, the molecule leaves the surface and cannot collide with other surface cubes. These assumptions hold for a small enough corrugation (sufficiently small $\alpha_{\text{max}}$).

For a molecule falling on a corrugated surface with an incident angle $\theta_\text{inc}$, the probability of hitting the surface depends on the horizontal coordinate $X'$. This probability is also proportional to the cosine of the effective incident angle $\theta_\text{inc}-\alpha$:
\begin{eqnarray}
\text{Prob}(\alpha)\propto \cos{(\theta_\text{inc}-\alpha)}&=&\cos{\theta_\text{inc}}\cos{\alpha}+\sin{\theta_\text{inc}}\sin{\alpha} \\
~&=&\frac{\cos{\theta_\text{inc}}}{\sqrt{1+\tan^2{\alpha}}}(1+\tan{\theta_\text{inc}}\tan{\alpha})~. \nonumber
\end{eqnarray}
Substituting Eq.\ (\ref{tan_alpha}) and normalizing gives the dependence on $X'$:
\begin{equation}
\text{Prob}(X')=\frac{2\pi}{Kc}\,\frac{1+r\tan{\theta_\text{inc}}\sin{(2\pi X'/c)}}{\sqrt{1+r^2\sin^2{(2\pi X'/c)}}}~, 
\end{equation}
where $r=2\pi a/c$, and
\begin{equation}
K=\int\limits_{0}^{2\pi}\frac{d\xi}{\sqrt{1+r^2\sin^2{\xi}}}~. 
\end{equation}

\subsection{Vibration of surface atoms - simple inclusion of phonons}
\label{vibration}

The hard cube model \cite{Goodman65,Logan66,Doll73} provides a simple way for including surface atom vibration into the collision process.
The collision of a molecule with a hard wall of infinite mass is replaced by a collision with a hard cube of a finite, but a large mass $M$ moving with velocity $U$. According to the model, the cube oscillates in the direction locally perpendicular to the surface. The reason is that for a flat and frictionless cube only the vertical velocity component can transfer energy to the impinging molecule. We slightly modify the original hard cube model and let the cube perform harmonic oscillations with a frequency $\omega_M$ (the spring constant of the oscillator is $M\omega_M^2$). The position of the face of the cube, in the unprimed coordinate system, is given by:
\begin{equation}
\label{Z'M}
Z_M(t)=A\sin{(\omega_M t+\varphi)}~, 
\end{equation}
and the velocity, therefore, is:
\begin{equation}
V_{Z,M}(t)=U(t)=\omega_M A\cos{(\omega_M t+\varphi)}~, 
\end{equation}
where $A$ and $\varphi$ are the amplitude and the phase of the oscillator, respectively. These $A$ and $\varphi$ will be chosen randomly, according to the distribution functions, as explained below.

When treating the collision process, we assume that the mass of the surface atom is much larger than the mass of the molecule, i.\ e.\ $M\gg 2m$, so that the collision does not change the velocity of the hard cube. Only the velocities of the molecule change, and its total energy in the laboratory coordinate frame may increase or decrease. If we assume that the oscillating hard cubes are always in thermal equilibrium, then the constant total energy of the harmonic oscillator, equal to $M\omega_M^2 A^2/2$, leads to the following probability distribution of the amplitudes $A$:
\begin{equation}
\label{surface_A_dist}
f(A)=\sqrt{\frac{M\omega_M^2}{2\pi k_BT_{\text{surf}}}}\exp{\left(-\frac{M\omega_M^2 A^2}{2k_BT_{\text{surf}}}\right)}~,
\end{equation}
where $T_{\text{surf}}$ is the temperature of the solid surface, and $k_B$ is the Boltzmann constant. The phase $\varphi$ of the oscillator is uniformly distributed between $0$ and $2\pi$.

The collision condition for a molecule falling on a surface cube can be expressed by
\begin{equation}
\label{condition_phonons}
Z(t)-Z_M(t)=\frac12 r_e |\cos{\theta(t)}|~,
\end{equation}
where $Z(t)$ is the time-dependent position of the center of mass of the colliding molecule, $Z_M(t)$ is given by Eq.\ (\ref{Z'M}), and $0.5 r_e \cos{\theta}$ is the $z$-component of the orientation vector $\mathbf{r}$ of the molecule. 

The translational velocity of the molecule is transformed {\it before} the collision according to $V_Z\to V_Z-U(t_{\text{col}})$, where $t_{\text{col}}$ is the moment of the collision. This is a usual Galilean transformation to the coordinate frame moving with the hard cube. Similarly, after the collision the translational velocity of the molecule is transformed back by $V_Z\to V_Z+U(t_{col})$ (the collision and the change of velocities occur instantaneously in our model). The rotational velocity remains unchanged under this Galilean transformation.

\subsection{The collision of the molecule with a frictionless surface}
\label{collision1}

The collision process in the local (unprimed) coordinate system can now be treated in the same way as in \cite{Khodorkovsky11}. Here we provide the final results, taken from the Appendix section of \cite{Khodorkovsky11} (after substituting $\mu=1$ for the homonuclear diatomic molecule.)

Here and below, the subscripts $i$ and $f$ denote the values of variables before and after a single collision, respectively. For a collision with a frictionless cube the parallel velocity component is always conserved: $V_{X,f}=V_{X,i}$. The colliding molecule can be therefore treated as if it had only a $Z$-component of its center-of-mass velocity. The final perpendicular velocity component is:
\begin{equation}
\label{Vf_3d+}
V_{Z,f}=\frac{-\cos^2{\theta}V_{Z,i}-2|\mathbf{v}_i|\sin{\theta}\left(\vec{e}_{v_i}\cdot\vec{e}_{v_f-v_i}\right)}{1+\sin^2{\theta}}~.
\end{equation}
Here $\vec{e}_{v_i}$ is the unit vector in the direction of the initial rotational velocity $\mathbf{v}_i$, and $\vec{e}_{v_f-v_i}$ is given by:
\begin{equation}
\label{unit_vec}
\vec{e}_{v_f-v_i}=-\frac{\vec{e}_{Z}-\cos{\theta}\,\mathbf{r}_i}{\sin{\theta}}~,
\end{equation}
where $\vec{e}_{Z}$ is the unit vector in the direction of $Z$.
The final rotational velocity $\mathbf{v}_f$ can be found using:
\begin{equation}
\label{vf_3d-}
\mathbf{v}_f=\mathbf{v}_i+\left|V_{Z,f}-V_{Z,i}\right|\sin{\theta}\,\vec{e}_{v_f-v_i}~.
\end{equation}

Before, after and between the collisions (if there are several), the motion of the molecule is in free space. The translational motion of the center of mass is:
\begin{equation}
\label{Zt}
Z(t)=Z_0+V_{Z,0}(t-t_0)~.
\end{equation}
The rotational motion of the molecule is a free rotation with a constant speed $v_0$. The plane of rotation is fixed and is perpendicular to the angular momentum vector, which is a constant of motion for free rotation. The orientation of the molecule as a function of time is:
\begin{equation}
\label{r_of_t}
\mathbf{r}(t)=\mathbf{r_0}\cos{[\omega_0(t-t_0)]}+\frac{r_e}{2}\frac{\mathbf{v_0}}{v_0}\sin{[\omega_0(t-t_0)]}~,
\end{equation}
where $v_0=|\mathbf{v_0}|$, $\omega_0$ is the angular rotational speed of the molecule, and $v_0=\omega_0r_e/2$. Taking the derivative, the velocity is:
\begin{equation}
\label{v_of_t}
\mathbf{v}(t)=-\omega_0\mathbf{r_0}\sin{[\omega_0(t-t_0)]}+\mathbf{v_0}\cos{[\omega_0(t-t_0)]}~.
\end{equation}

\section{Interaction of the molecule with an ultrashort laser pulse}
\label{laser}

Here we briefly summarize the results of the classical model describing the interaction of the diatomic rigid molecule with a nonresonant ultrashort laser pulse, in the impulsive approximation. A more detailed description may be found in \cite{Khodorkovsky}.

The potential energy of the laser pulse interacting with the induced molecular dipole moment is given by:
\begin{equation}
\label{V}
V(\theta',\varphi',t)=-\frac{1}{4}\mathcal{E}^2(t)\left(\Delta\alpha\cos^2{\beta'}+\alpha_{\perp}\right)~,
\end{equation}
where $\Delta\alpha=\alpha_{\parallel}-\alpha_{\perp}$ is the difference between the polarizability along the molecular axis and the one perpendicular to it, $\mathcal{E}(t)$ is the envelope of the electric field of the {\it linearly polarized} laser pulse, and $\beta'=\beta'(\theta',\varphi')$ is the angle between the molecular axis and the direction of polarization of the pulse. We assume that the pulse duration is very short compared to the rotational period, so that the pulse can be described in the impulsive ($\delta$-kick) approximation. We define the dimensionless pulse strength $P$, as
\begin{equation}
\label{P}
P=\frac{\Delta\alpha}{4\hbar}\int_{-\infty}^{\infty}\mathcal{E}^2(t)dt~.
\end{equation}
Planck's constant $\hbar$ appears here and in the remainder of the paper only for the reason of unit convention. Under this convention, the quantity $\hbar P$ is the typical value of angular momentum transferred by the pulse to the molecule.

We consider the action of a pulse linearly polarized along some arbitrary unit vector $\mathbf{p}$, and determine the vector of the resulting velocity change $\boldsymbol{\Delta}\mathbf{v}'$ for a molecule oriented along some direction $\mathbf{r_0}'$. The final result is:
\begin{equation}
\label{delta_v_vec}
\boldsymbol{\Delta}\mathbf{v}'=\frac{2\hbar P}{I}\cos{\beta_0'}\left(\frac{r_e}{2}\mathbf{p}-\cos{\beta_0'}\mathbf{r_0}'\right)~,
\end{equation}
where $I=mr_e^2/2$ is the moment of inertia of the molecule.

\section{A surface as a ``tennis-racquet'' selector of molecules: a correlation between the scattering angle and the sense of rotation}
\label{results1}

\begin{figure}
\centering
\includegraphics[width=1\textwidth]{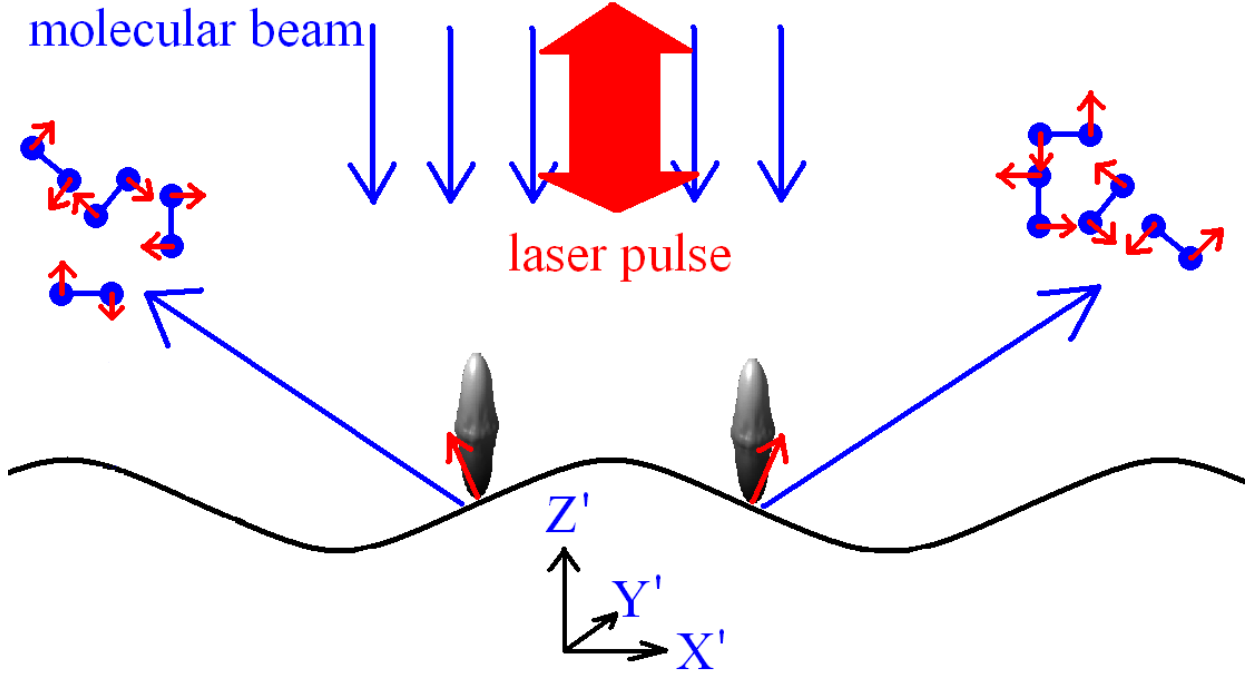}
\caption{(Color online) The suggested experiment, in which the molecular beam falling on the corrugated surface at an angle $\theta_{\text{inc}}=0^{\circ}$ (blue thin arrows) is irradiated by an ultrashort laser pulse (thick red arrow), polarized as shown. This laser pulse aligns the molecules and produces a time-averaged angular distribution of them, as depicted by the gray cigars. These aligned molecules fall on two different slopes of the corrugated surface. They receive a differently directed ``kick'' on the two slopes (small red arrows), and as a result, they scatter to different angles, with a different sense of rotation, as shown.}
\label{corrug_pic_correl}
\end{figure}

In our previous publication \cite{Khodorkovsky11} we suggested a way to produce diatomic molecules, such as $\text{N}_2$, with a preferred sense of rotation using an ultrashort laser pulse and an event of molecular scattering from a {\it flat} solid surface [such as $\text{Ag}(111)$]. We suggested to shine a femtosecond laser pulse polarized at $+45^{\circ}/-45^{\circ}$ to the surface normal on molecules in a molecular beam. As a result, the flying molecules will be aligned (on average) along the direction of polarization of the pulse. After hitting the flat surface, the molecules will be ``kicked'' by it and receive a preferential rotational velocity of clockwise/counter-clockwise sense in the plane containing the polarization vector of the laser pulse and the surface normal.

\begin{figure}
\centering
\includegraphics[width=1\textwidth]{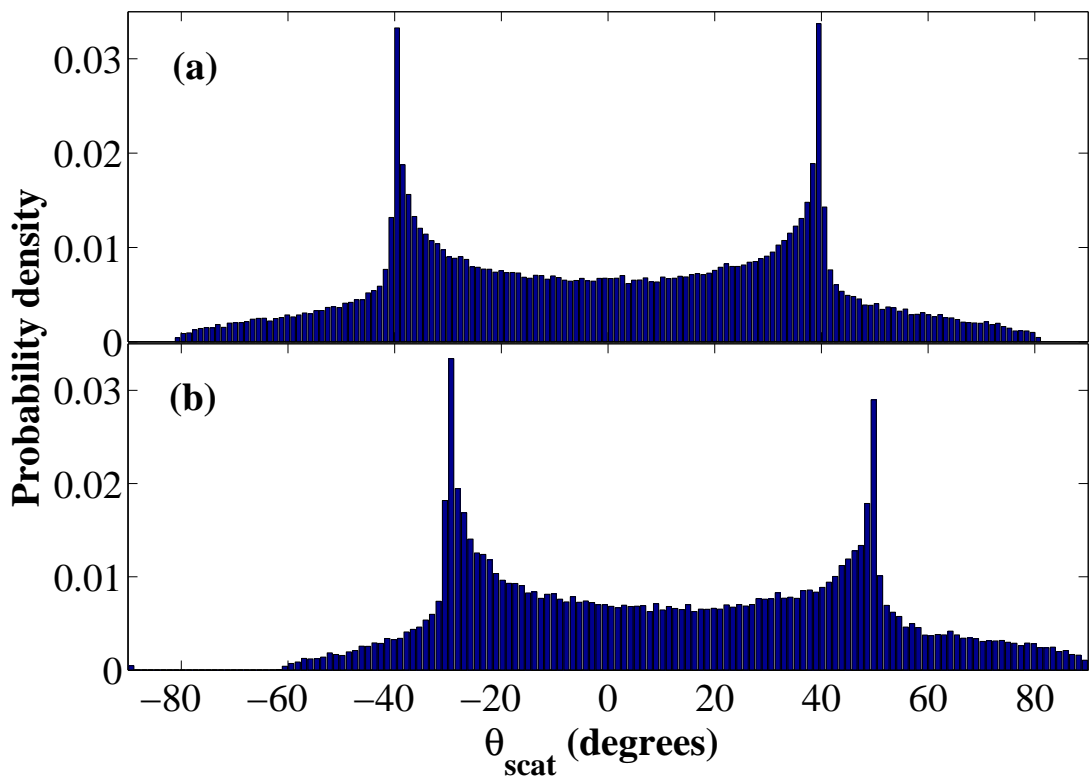}
\caption{Probability distributions of the scattering angles of non-rotating ($T_{\text{rot}}=0$) nitrogen molecules scattered from a frozen corrugated surface ($T_{\text{surf}}=0$, $\alpha_{\text{max}}=20^{\circ}$), with an incident angle of (a) $\theta_{\text{inc}}=0^{\circ}$ and (b) $\theta_{\text{inc}}=-10^{\circ}$. The corrugation rainbows can be seen at expected angles of $\theta_{\text{rainbow}}=-\theta_{\text{inc}}\pm 2\alpha_{\text{max}}$. In (b) the left peak is higher than the right peak, because the incident molecules have a higher probability to fall on the positive slope of the corrugated surface, and to scatter to negative angles. In this Figure, the results of a Monte Carlo simulation of $10^5$ molecules are shown.}
\label{inc0_10_no_pulse}
\end{figure}

Our current suggestion utilizes laser-aligned molecules that are scattered from a {\it corrugated} surface, such as $\text{LiF}(001)$, or $\text{Pt}(211)$. The idea is based on the following consideration. Consider diatomic molecules that are ``shot'' towards the corrugated surface with $\theta_{\text{inc}}=0^{\circ}$, see Fig.\ \ref{corrug_pic_correl}. Part of these molecules will fall on a corrugation with {\it negative} slope and will be scattered mainly to {\it positive} scattering angles $\theta_{\text{scat}}$, while the other part will fall on a corrugation with {\it positive} slope and will be scattered mainly to {\it negative} scattering angles. Now consider irradiating these molecules with an ultrashort laser pulse polarized along the surface normal, at $0^{\circ}$, before they hit the surface. The molecules will now be aligned on average at $0^{\circ}$ when hitting the surface. The molecules hitting the {\it negative} slope of the corrugation will scatter to {\it positive} angles and receive a ``kick'' providing them a {\it counter-clockwise} rotation, while the molecules hitting the {\it positive} slope of the corrugation will scatter to {\it negative} angles and receive a ``kick'' providing them a {\it clockwise} rotation. Therefore, a {\it correlation} between the angle of scattering and the direction of rotation should be present in the distribution of the scattered molecules.

\begin{figure}
\centering
\includegraphics[width=1\textwidth]{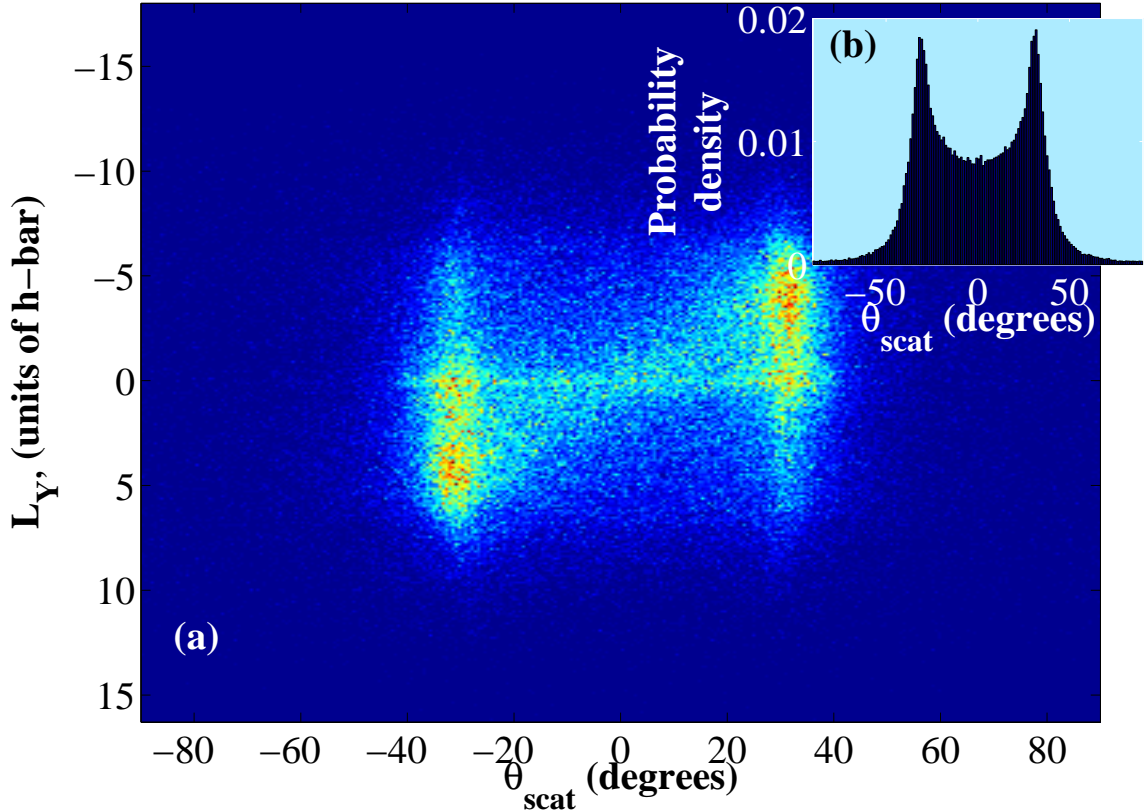}
\caption{(Color online) A two-dimensional probability distribution of the scattered molecules. The horizontal axis represents the scattering angle, and the vertical axis is for the angular momentum component along the $Y'$-direction, in units of $\hbar$. Hot (red) colors represent high probability, while cold (blue) colors represent low probability. Before the scattering, the molecules were ``kicked'' by a laser pulse polarized at $0^{\circ}$, along the $Z'$-axis, with a pulse strength of $P=10$. The angle of incidence is $\theta_{\text{inc}}=0^{\circ}$. A clear correlation between the scattering angle and the sense of rotation is seen. In (b), a distribution of the scattering angles of the molecules is shown. The incident nitrogen molecules ($m=14\,\text{a.u.}$) have a speed of $V_0=300\,\text{m/s}$ and a rotational temperature of $T_{\text{rot}}=5\,\text{K}$. The corrugated surface is composed of platinum atoms ($M=195\,\text{a.u.}$) vibrating at a frequency of $\omega_M=10^{13}\,\text{Hz}$, while the amplitudes of vibration are distributed according to Eq.\ (\ref{surface_A_dist}) with $T_{\text{surf}}=300\,\text{K}$. The corrugation strength of the surface is $\alpha_{\text{max}}=20^{\circ}$. In this Figure and in Figs.\ \ref{inc10_with_pulse_0angpol} and \ref{inc0_with_pulse_90angpol} the results of Monte Carlo simulations of $2\times 10^5$ molecules are shown.}
\label{inc0_with_pulse_0angpol}
\end{figure}

When atoms or molecules are scattered from a corrugated surface, rainbows might be seen in the scattering angle distribution of the scattered atoms/molecules, see the review \cite{Kleyn91} and the experiment in \cite{Kondo05}. The angles corresponding to the rainbow scattering are determined by the points of inflection of the corrugation function. For a frozen (zero temperature) hard-wall surface with a corrugation given by Eq.\ (\ref{corrugation}), the rainbow scattering angles are easily shown to be equal to \cite{Kleyn91}
\begin{equation}
\label{rainbow}
\theta_{\text{rainbow}}=-\theta_{\text{inc}}\pm 2\alpha_{\text{max}}~. 
\end{equation}

The simulation results discussed below were obtained using the models of Sections \ref{model1} and \ref{laser} and a Monte Carlo averaging.
In Fig.\ \ref{inc0_10_no_pulse} we plot the angular distribution of non-rotating molecules scattered from a frozen corrugated surface ($T_{\text{surf}}=0$ and $\alpha_{\text{max}}=20^{\circ}$). In panel (a), the incident angle is $\theta_{\text{inc}}=0^{\circ}$, and in panel (b) $\theta_{\text{inc}}=-10^{\circ}$. In (a) the peaks correspond to Eq.\ (\ref{rainbow}) and are symmetric, because the incident molecules hit with equal probability the positive and the negative slopes of the surface corrugation. In (b) the peaks also correspond to Eq.\ (\ref{rainbow}) and the left peak is higher than the right peak. The reason is that for negative incident angles the probability of hitting the positive slope of the corrugation and scattering towards negative angles is higher than the same probability for the negative slope.

\begin{figure}
\centering
\includegraphics[width=1\textwidth]{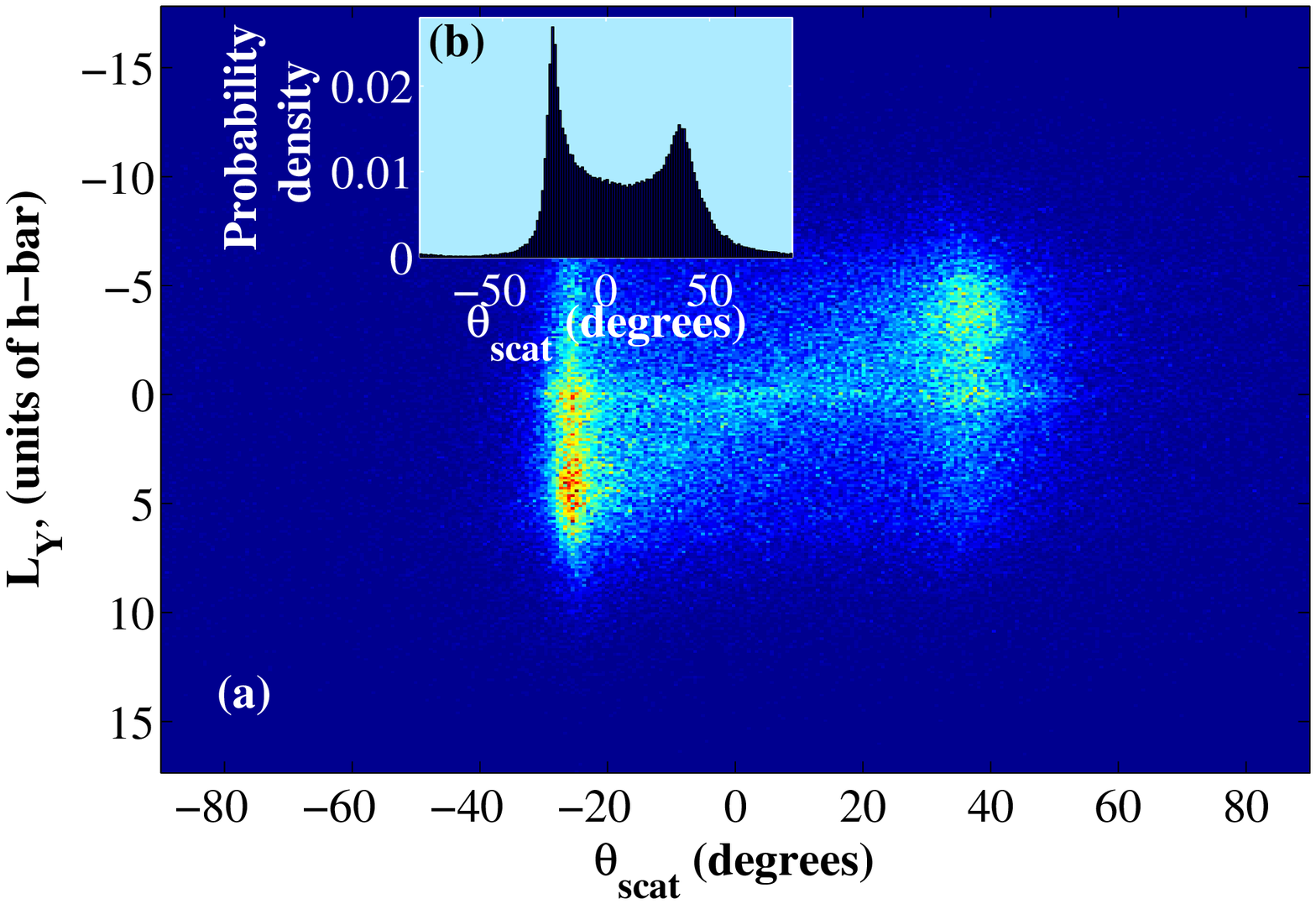}
\caption{(Color online) The same distribution as in Fig.\ \ref{inc0_with_pulse_0angpol} is plotted, except the incident angle that is $\theta_{\text{inc}}=-10^{\circ}$ here. The slight rotation of the molecular beam (or the surface) leads to a change of the relative intensity of the two rainbow peaks.}
\label{inc10_with_pulse_0angpol}
\end{figure}

\begin{figure}
\centering
\includegraphics[width=1\textwidth]{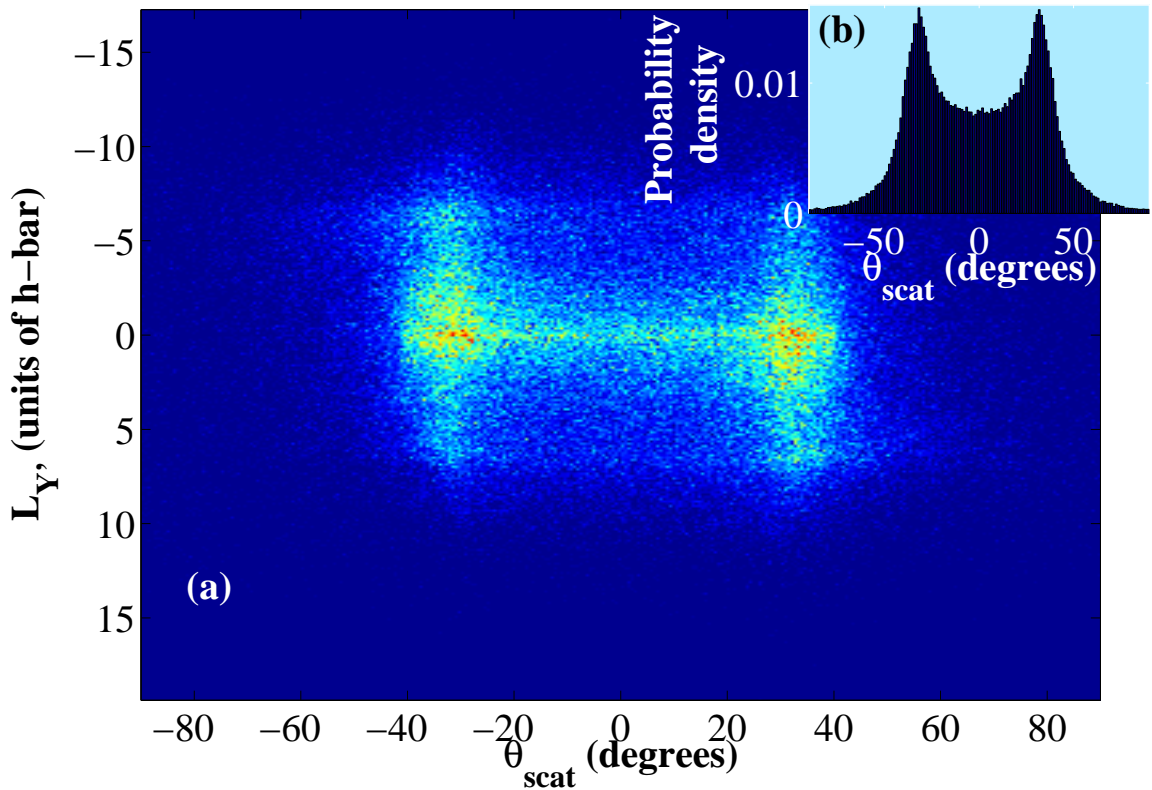}
\caption{(Color online) The same distribution as in Fig.\ \ref{inc0_with_pulse_0angpol} is plotted, except the direction of the polarization of the laser pulse, that is $90^{\circ}$ here, or along the $X'$-axis. Now, the incident molecules fall almost parallel to the weakly corrugated surface. This leads to double-collisions of the molecules, where the second collision almost cancels out the rotational excitation of the first collision. This still leaves some correlation, which is, however, much weaker than in Fig.\ \ref{inc0_with_pulse_0angpol}.}
\label{inc0_with_pulse_90angpol}
\end{figure}

In Figs.\ \ref{inc0_with_pulse_0angpol}, \ref{inc10_with_pulse_0angpol} and \ref{inc0_with_pulse_90angpol}, we plot the results for a molecular beam of nitrogen molecules falling on the corrugated surface ($\alpha_{\text{max}}=20^{\circ}$) at several incident angles and at a constant speed of $V_0=300\,\text{m/s}$. The molecules in the beam have a rotational temperature of $T_{\text{rot}}=5\,\text{K}$, typical for molecular beam experiments. The surface is composed of cubes of mass $M=195\,\text{a.u.}$, corresponding to platinum atoms, with oscillation frequency of $\omega_M=10^{13}\,\text{Hz}$. The amplitudes of the vibrating cubes are distributed according to Eq.\ (\ref{surface_A_dist}), with a temperature of $T_{\text{surf}}=300\,\text{K}$.

In Fig.\ \ref{inc0_with_pulse_0angpol}, we ``shoot'' the molecules towards the corrugated surface at normal incidence ($\theta_{\text{inc}}=0^{\circ}$), ``kick'' them with an ultrashort laser pulse polarized along the same vertical direction, and let them scatter from the corrugated surface
In Fig.\ \ref{inc0_with_pulse_0angpol}(a) we plot a two-dimensional histogram, where the vertical axis represents the angular momentum component $L_{Y'}$ perpendicular to the plane of the expected molecular rotation, and the horizontal axis represents the scattering angle $\theta_{\text{scat}}$. A clear correlation can be seen between the angle of scattering and the sign of the angular momentum component. Molecules with positive $L_{Y'}$ are scattered to negative angles, as expected. 

The distribution of scattering angles [which is also shown in the two-dimensional plot in Fig.\ \ref{inc0_with_pulse_0angpol}(b)] exhibits a double rainbow form, as in Fig.\ \ref{inc0_10_no_pulse}. However, the location of the peaks is at smaller values of $|\theta_{\text{scat}}|$ than the values given by Eq.\ (\ref{rainbow}). This happens for two reasons. First, the surface has a non-zero temperature, and the vibrating surface cubes transfer translational velocity to the molecules in the $Z$-direction (the corrugation is rather small and all the cubes vibrate mainly along the $Z$ direction). This transfer of velocity moves the scattering angle closer to the surface normal. Second, the laser pulse provides a high rotational energy to the molecules. During the collision this high rotational energy is partly transferred to a translational energy. Because the $V_X$ velocity component is conserved, most of the energy goes to the $V_Z$ component, which, because the surface corrugation is small, again, moves the scattering angle closer to the surface normal.

The same kind of two-dimensional distribution for $\theta_{\text{inc}}=-10^{\circ}$ is shown in Fig.\ \ref{inc10_with_pulse_0angpol}. Here the two rainbow peaks move to different angles, as seen qualitatively from Eq.\ (\ref{rainbow}), and their relative intensity changes, as explained above. This suggests a way to control the intensity of molecular beams that are scattered from a solid surface with a preferred sense of rotation.

Going back to the case of $\theta_{\text{inc}}=0^{\circ}$ and changing the polarization of the laser to $90^{\circ}$, we obtain Fig.\ \ref{inc0_with_pulse_90angpol}. The correlation between $L_{Y'}$ and $\theta_{\text{scat}}$ is now inverted, as could be expected. In addition, the effect now is much weaker than for the case of $0^{\circ}$ polarization of the laser pulse, in Fig.\ \ref{inc0_with_pulse_0angpol}. The reason is that now the molecules that are falling on the surface are aligned parallel to the surface. Because the surface is almost flat ($\alpha_{\text{max}}=20^{\circ}$), these molecules can collide twice with the surface, when the second collision almost cancels out the rotational excitation caused by the first collision.

\section{A model of molecular scattering from a surface with friction}
\label{model2}

In this Section we develop a model and analyze the scattering of a molecule from a surface that is ``corrugated'' at a length scale much smaller than the size of the molecule. This ``corrugation'' is treated as a friction force acting parallel to the flat surface. The prototype for our model is the model that was used in order to describe qualitatively some effects in the scattering of nitrogen molecules from the Ag(111) surface \cite{Sitz88}. This silver surface is practically flat, however, the measured effects in \cite{Sitz88} suggest that in-plane tangential forces are present during the collision process. These in plane tangential forces can be created by different physical mechanisms, such as: tangentially directed phonons, electron-hole pair creation, etc. 

Similar to Sec.\ \ref{model1}, we extend our model introduced in reference \cite{Khodorkovsky11}. In this Section, we add a friction force between the colliding atom and the flat surface, in a similar way to \cite{Sitz88}. The friction force is exerted parallel to the surface, its magnitude is proportional to the local velocity of the colliding atom and its direction is opposite to this local velocity direction. In subsection \ref{collision2_2d}, the model is developed for a frozen surface, and for a two-dimensional planar rotation of the molecule. In subsection \ref{collision2_3d}, the model is extended to a three-dimensional rotation. The extension to a finite-temperature surface is done as in subsection \ref{vibration}.

\subsection{Definition of the coordinates}
\label{coordinates2}

\begin{figure}
\centering
\includegraphics[width=1\textwidth]{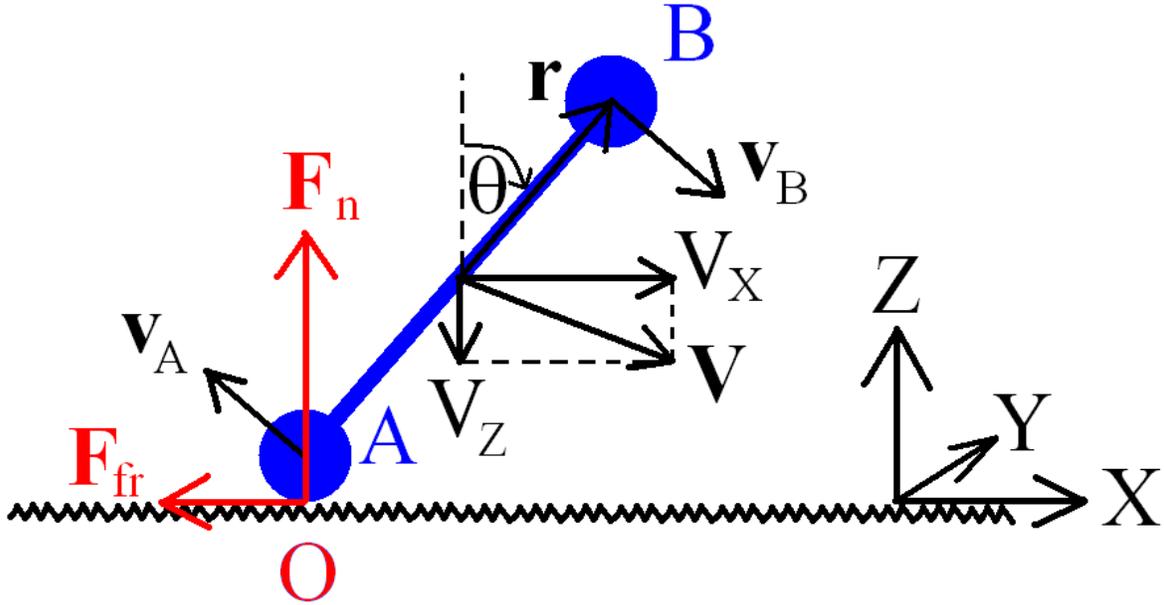}
\caption{(Color online) Diatomic molecule (a dumbbell) colliding with a flat, rough surface. The molecule has translational velocity $\mathbf{V}$, with $X$ and $Z$ components $V_X$ and $V_Z$, respectively. The rotational velocity of atoms $A$ and $B$ is $\mathbf{v}_A$ and $\mathbf{v}_B$, respectively, where $|\mathbf{v}_A|=|\mathbf{v}_B|=v$. The position of the center of mass is given by the coordinates $X$, $Y$ and $Z$, and the orientation vector of the molecule is $\mathbf{r}$. The forces acting on the atom $A$ at the point of contact $O$ are shown in red: $\mathbf{F}_{\text{n}}$ is the normal force, and $\mathbf{F}_{\text{fr}}$ is the friction force, given by Eq.\ (\ref{force}) for the two dimensional case, and by Eq.\ (\ref{force3d}) for the three-dimensional case.}
\label{molecule_surface_friction}
\end{figure}

The model we use in this and the following Sections treats the surface as a flat surface with friction, rather than a corrugated wall (later in subsection \ref{results2_rot_3d}, the surface is composed of vibrating hard flat cubes). The variables and the parameters are similar to those of Sec.\ \ref{model1}, with few exceptions. The $(X,Y,Z)$ and the $(X',Y',Z')$ coordinate systems are now identical one to another. The coordinate system from this point on is $(X,Y,Z)$, as shown in Fig.\ \ref{molecule_surface_friction}. Because the frictional forces can act also in the $Y$-direction, the center-of-mass velocity should include now three components: $\mathbf{V}=(V_X,V_Y,V_Z)$. The molecular orientation is $\mathbf{r}$, and the rotational velocity is $\mathbf{v}$. The mass of the molecule is $2m$. The incident angle is $\theta_{\text{inc}}=\arctan{(V_{X,0}/V_{Z,0})}$ (initially $V_Y$ is always chosen as zero). The point of collision is denoted by $O$. The colliding atom is denoted by $A$, and the other atom is denoted by $B$. It is assumed that the special case when both atoms collide simultaneously with the surface never occurs.

\subsection{A two-dimensional model of molecular scattering from a surface with friction}
\label{collision2_2d}

If the translational and rotational motion of the molecule is restricted to the $XZ$-plane, the $Y$ coordinate and the $V_Y$ velocity component are always equal to zero. The orientation of the molecule in the $XZ$-plane can now be defined by a single angle $\theta$ that increases in a clockwise direction from the direction of the $Z$-axis and ranges from $0$ to $2\pi$. The rotational velocity is now defined by a single variable $v=|\mathbf{v}_A|=|\mathbf{v}_B|$, equal to $\omega r_e/2$, where $\omega$ is the angular velocity of the molecule.

\subsubsection{Friction force}

At the moment of the collision, two forces act on the colliding atom $A$: a normal force $\mathbf{F_{\text{n}}}$ and a friction force $\mathbf{F_{\text{fr}}}$, see Fig.\ \ref{molecule_surface_friction}. We assume that $\mathbf{F_{\text{n}}}$ is a conservative force, that its functional form is not known. However, we assume that $\mathbf{F_{\text{fr}}}$ is a nonconservative force that has a known functional form, given below. We also assume that the time interval during which the forces are applied is very short compared to the time it takes to the variables $X$, $Z$ and $\theta$ to change significantly, such that the forces are practically instantaneous.

As explained above, the friction force is given by:
\begin{equation}
\label{force}
\mathbf{F_{\text{fr}}}=-2m\gamma \left[(\mathbf{V}+\mathbf{v_A})\cdot \mathbf{e_X}\right]\mathbf{e_X} \,\delta(t)~, 
\end{equation}
where $\gamma$ is the friction coefficient, $\mathbf{e_X}$ is a unit vector in the $X$ direction, and $\delta(t)$ is the Dirac delta function in the time variable. The time $t=0$ denotes the moment the molecule arrives to its turning point during the collision with the surface. It is assumed here and below, that $\gamma\ll 1$. The meaning of this assumption will be clear in the next paragraph.

\subsubsection{Solution of equations of motion and conservation of angular momentum}

{\bf Momentum change in the $X$-direction.} In the following derivations, we assume that the orientation $\theta$ of the molecule is in the range between $0$ and $\pi/2$. Later, we extend the final results for any angle. Newton's equation of motion for the center of mass motion in the $X$-direction is:
\begin{equation}
2m\frac{dV_X}{dt}=-2m\gamma\left[V_X(t)-v(t)\cos{\theta(t)}\right]\delta(t)~. 
\end{equation}
After integrating over the time of the collision, we obtain:
\begin{equation}
V_X^+-V_X^-=-\gamma\left[V_X(0)-v(0)\cos{\theta}\right]~, 
\end{equation}
where the superscripts $-$ and $+$ denote the values of the variables immediately before and after the collision, respectively. We assume that during the collision the position and the orientation of the molecule do not change, while the velocities do change. We assume that the velocities at the turning point $V_X(0)$ and $v(0)$ are equal to the velocities before the collision $V_X^-$ and $v^-$, respectively. This assumption is valid, because our basic assumption is that $\gamma\ll 1$, and this means that the change in $V_X$ as the result of the friction force is small, compared to $V_X$ itself. We obtain:
\begin{equation}
\label{Vx+_full}
V_X^+=V_X^--\gamma\left(V_X^--v^-|\cos{\theta}|\right)~, 
\end{equation}
where the absolute value makes the expression valid also for values of $\theta$ higher than $\pi/2$. Defining the effective speed of the colliding atom before the collision as:
\begin{equation}
V_{X,\text{eff}}^-=V_X^--|\cos{\theta}|v^-~, 
\end{equation}
Eq.\ (\ref{Vx+_full}) can be written as:
\begin{equation}
\label{Vx+}
V_X^+=V_X^--\gamma V_{X,\text{eff}}^-~.  
\end{equation}

{\bf Angular momentum conservation with respect to the origin $O$.} Taking the point $O$ to be the origin (see Fig.\ \ref{molecule_surface_friction}), it can be seen that the lever arms of the forces $\mathbf{F_{\text{n}}}$ and $\mathbf{F_{\text{fr}}}$ are zero. This leads to the conclusion that the collision process preserves the angular momentum of the molecule, measured with respect to the origin $O$. In addition, the angular momentum of atom $A$ is zero, so the conservation law is:
\begin{equation}
L_B^+=L_B^-~, 
\end{equation}
or, explicitly:
\begin{equation}
\label{eq1}
\cos{\theta}\,V_X^+-\sin{\theta}\,V_Z^++v^+=\cos{\theta}\,V_X^--\sin{\theta}\,V_Z^-+v^-~, 
\end{equation}
similar to Eq.\ (4) in \cite{Khodorkovsky11}.

{\bf Energy change.} The total molecular energy is the sum of the translational energy of the center of mass and the rotational energy with respect to the center of mass, and is given by:
\begin{equation}
\label{energy}
E=\frac{1}{2}(2m)(V_X^2+V_Z^2)+\frac{1}{2}(2m)v^2~. 
\end{equation}
The rate of change of the energy during the collision, or the power, is given by:
\begin{equation}
\label{energy_change}
\frac{dE}{dt}=(\mathbf{F_{\text{n}}}+\mathbf{F_{\text{fr}}})\cdot (\mathbf{V}+\mathbf{v_A})~.
\end{equation}
Integrating over the collision time, noticing that $\mathbf{F_{\text{n}}}$ is a conservative force that does not contribute to the change of energy and using Eq.\ (\ref{force}), we obtain:
\begin{equation}
\label{energy_change2}
E^+-E^-=-2m\gamma\left[V_X(0)-v(0)\cos{\theta}\right]^2~. 
\end{equation}
Using Eq.\ (\ref{energy}), assuming that the velocities at the turning point are equal to the velocities before the collision and performing some rearrangements, we arrive at:
\begin{equation}
\label{eq2}
(V_X^+)^2+(V_Z^+)^2+(v^+)^2=(1-2\gamma)(V_X^-)^2+(V_Z^-)^2+4\gamma v^-V_X^-\cos{\theta}+(1-2\gamma\cos^2{\theta})(v^-)^2~. 
\end{equation}

\subsubsection{The final velocities}

The velocity $V_X^+$ is given by Eq.\ (\ref{Vx+}). Using Eqs.\ (\ref{Vx+}), (\ref{eq1}) and (\ref{eq2}) the values of $V_Z^+$ and $v^+$ can be found using tedious but straightforward algebra. We solve a quadratic equation, and the sign before the square root is chosen such that the following equations for $\gamma=0$ are consistent with Eqs.\ (\ref{Vf_3d+}) and (\ref{vf_3d-}). At the end, we generalize the final results for the orientation angle $\theta$ in the range from $0$ to $2\pi$, by adding an absolute value [as in Eq.\ (\ref{Vx+_full})], and a correct sign, where necessary. We also neglect terms proportional to $\gamma^2$, while leaving terms proportional to $\gamma$, for consistency with the assumption $\gamma\ll 1$. The final result is:
\begin{equation}
\label{Vz+}
V_Z^+=-\frac{\cos^2{\theta}\,V_Z^-+2\,\text{sgn}(\tan{\theta})|\sin{\theta}|v^-+\gamma\sin{2\theta}\,V_{X,\text{eff}}^-}{1+\sin^2{\theta}}~. 
\end{equation}
Using Eqs.\ (\ref{Vx+}) and (\ref{eq1}), the result for $v^+$, expressed using $V_Z^+$ is:
\begin{equation}
\label{v+_pre}
v^+=v^-+\text{sgn}(\tan{\theta})|\sin{\theta}|(V_Z^+-V_Z^-)+\gamma|\cos{\theta}|V_{X,\text{eff}}^-~, 
\end{equation}
or, after substituting Eq.\ (\ref{Vz+}):
\begin{equation}
\label{v+}
v^+=\frac{\cos^2{\theta}\,v^--2\,\text{sgn}(\tan{\theta})|\sin{\theta}|V_Z^-+\gamma|\cos^3{\theta}|\,V_{X,\text{eff}}^-}{1+\sin^2{\theta}}~. 
\end{equation}
In the equations above, $\text{sgn}(a)$ denotes the signum function, defined as $\text{sgn}(a)=a/|a|$.

The free motion of the molecule before, after, and in between the collisions is according to Eqs.\ (\ref{Zt}), (\ref{r_of_t}), (\ref{v_of_t}). Equations (\ref{r_of_t}) and (\ref{v_of_t}) for a two-dimensional rotation reduce to:
\begin{equation}
\theta(t)=\theta_0+\omega_0 (t-t_0)~. 
\end{equation}
The condition for collision of the molecule with the surface is given by Eq.\ (\ref{condition_phonons}), where $Z_M(t)=0$ for a frozen surface.

\subsection{A three-dimensional model of molecular scattering from a surface with friction}
\label{collision2_3d}

Now we extend the model to the case of a three-dimensional rotation. By our convention (see Fig.\ \ref{molecule_surface_friction}), at the moment of collision the atom touching the surface is denoted by $A$, while the other atom is denoted by $B$. It follows that at the moment of collision the orientation of atom $A$ is given by $\mathbf{r_A=-r}$, if the $z$-component of the orientation vector $\mathbf{r}$ is positive, and it is given by $\mathbf{r_A=r}$, if the $z$-component of the orientation vector $\mathbf{r}$ is negative. 

The rotational velocity of the molecule is now given by the vector $\mathbf{v}$, which is changing with time, as opposed to the scalar quantity $v=|\mathbf{v}|$, which is constant for a freely rotating molecule. The components of the velocity $\mathbf{v}$ are given by $(v_x,v_y,v_z)$, where the $(x,y,z)$ coordinate system is fixed to the center of mass of the molecule and is parallel to the $(X,Y,Z)$ system.

In the three-dimensional rotation case, the velocity component $V_Y$, that is initially zero, can become nonzero, because of the friction forces acting in the $Y$-direction.

\subsubsection{Friction force}

The friction force is now proportional to the local velocity component of atom $A$ parallel to the surface, and is given by:
\begin{equation}
\label{force3d}
\mathbf{F_{\text{fr}}}=-2m\gamma \left[\mathbf{e_X}(\mathbf{V}+\mathbf{v_A})\cdot \mathbf{e_X} + \mathbf{e_Y}(\mathbf{V}+\mathbf{v_A})\cdot \mathbf{e_Y}\right] \,\delta(t)~. 
\end{equation}

\subsubsection{Solution of equations of motion and conservation of angular momentum}

Similar to Eq.\ (\ref{Vx+_full}) we obtain for the velocity components $V_X^+$ and $V_Y^+$ after the collision:
\begin{equation}
\label{Vx+_full_3d}
V_X^+=V_X^--\gamma\left(V_X^-+v_{A,x}^-\right)~, 
\end{equation}
and
\begin{equation}
\label{Vy+_full_3d}
V_Y^+=V_Y^--\gamma\left(V_Y^-+v_{A,y}^-\right)~.
\end{equation}

Finding the energy change according to Eq.\ (\ref{energy_change}), we obtain, similar to Eq.\ (\ref{energy_change2}):
\begin{equation}
\label{energy_change2_3d}
E^+-E^-=-2m\gamma\left[(V_X^-+v_{A,x}^-)^2+(V_Y^-+v_{A,y}^-)^2\right]~. 
\end{equation}
In the last equation $E$ denotes the total molecular energy, equal to:
\begin{equation}
E=\frac{1}{2}(2m)\left(V_X^2+V_Y^2+V_Z^2+v_{A,x}^2+v_{A,y}^2+v_{A,z}^2\right)~. 
\end{equation}

The equation for angular momentum conservation, similar to Eq.\ (\ref{eq1}), is now expressed in vector form:
\begin{equation}
\mathbf{r}_B\times\left(\mathbf{V}^+-\mathbf{V}^-+\mathbf{v}_B^+-\mathbf{v}_B^-\right)=0~, 
\end{equation}
or, changing the orientation and rotational velocity vectors from the non-colliding atom $B$ to the colliding atom $A$:
\begin{equation}
\label{momentum_cons_3d}
\mathbf{r}_A\times\left(\mathbf{V}^--\mathbf{V}^++\mathbf{v}_A^+-\mathbf{v}_A^-\right)=0~. 
\end{equation}

\subsubsection{The final velocities}

Solving together Eqs.\ (\ref{Vx+_full_3d}), (\ref{Vy+_full_3d}), (\ref{energy_change2_3d}) and (\ref{momentum_cons_3d}) (Mathematica 7 software was used), and neglecting terms proportional to $\gamma^2$ or a higher power, we obtain the following expression for $V_Z^+$:
\begin{equation}
\label{Vz_3d_fric}
V_Z^+=\frac{\sin^2{\theta}(V_Z^--v_{A,z}^-)+0.5\sin{2\theta}\left( v_{\perp}^--\gamma V_{\perp}^-\right)-\text{sgn}(R^-)\left( \sqrt{\Delta}-2\gamma\sin{\theta}\, V_{\perp}^- R^-/\sqrt{\Delta}\right)}{1+\sin^2{\theta}}~, 
\end{equation}
where
\begin{equation}
v_{\perp}=v_{A,x}\cos{\varphi}+v_{A,y}\sin{\varphi}~, 
\end{equation}
\begin{equation}
V_{\perp}=(V_X+v_{A,x})\cos{\varphi}+(V_Y+v_{A,y})\sin{\varphi}~, 
\end{equation}
\begin{equation}
R=\sin{\theta}\,v_{\perp}-\frac{1}{2}\cos{\theta}(V_Z-v_{A,z})~, 
\end{equation}
and
\begin{equation}
\Delta=2R^2+\frac{1}{2}(1+\sin^2{\theta})(V_Z+v_{A,z})^2~. 
\end{equation}
Eq.\ (\ref{Vz_3d_fric}) reduces to Eq.\ (\ref{Vz+}) in the special case of rotation in the $XZ$ plane. In addition, we checked numerically that Eq.\ (\ref{Vz_3d_fric}) reduces to Eq.\ (\ref{Vf_3d+}) for $\gamma=0$.
From a vector diagram drawn according to Eq.\ (\ref{momentum_cons_3d}), and from the fact that $\mathbf{v}_A$ is always perpendicular to $\mathbf{r}_A$, we find an expression for $\mathbf{v}_A^+$, similar to Eq.\ (\ref{vf_3d-}):
\begin{equation}
\mathbf{v}_A^+=\mathbf{v}_A^-+\left|\mathbf{V}^+-\mathbf{V}^-\right|\sin{\alpha}\,\vec{e}_{\mathbf{v}_A^+-\mathbf{v}_A^-}~, 
\end{equation}
where
\begin{equation}
\sin{\alpha}=\sqrt{1-\left(\mathbf{r_A}\cdot \vec{e}\right)^2}~, 
\end{equation}
\begin{equation}
\vec{e}_{\mathbf{v}_A^+-\mathbf{v}_A^-}=-\frac{\vec{e}-\left(\mathbf{r}_A\cdot \vec{e}\right)\mathbf{r}_A}{\sin{\alpha}} ~,
\end{equation}
and
\begin{equation}
\vec{e}=\frac{\mathbf{V}^--\mathbf{V}^+}{\left|\mathbf{V}^--\mathbf{V}^+\right|} ~.
\end{equation}
The last equations reduce to equations from Sec.\ \ref{collision1} if $\vec{e}$ is replaced by $\vec{e}_Z$. This can be done in the case of scattering from a frictionless surface, because in such a collision the $V_X$ and $V_Y$ components are conserved, and the collision can be treated as a collision perpendicular to the surface, leading to $\vec{e}=\vec{e}_Z$.

\begin{figure}
\centering
\includegraphics[width=1\textwidth]{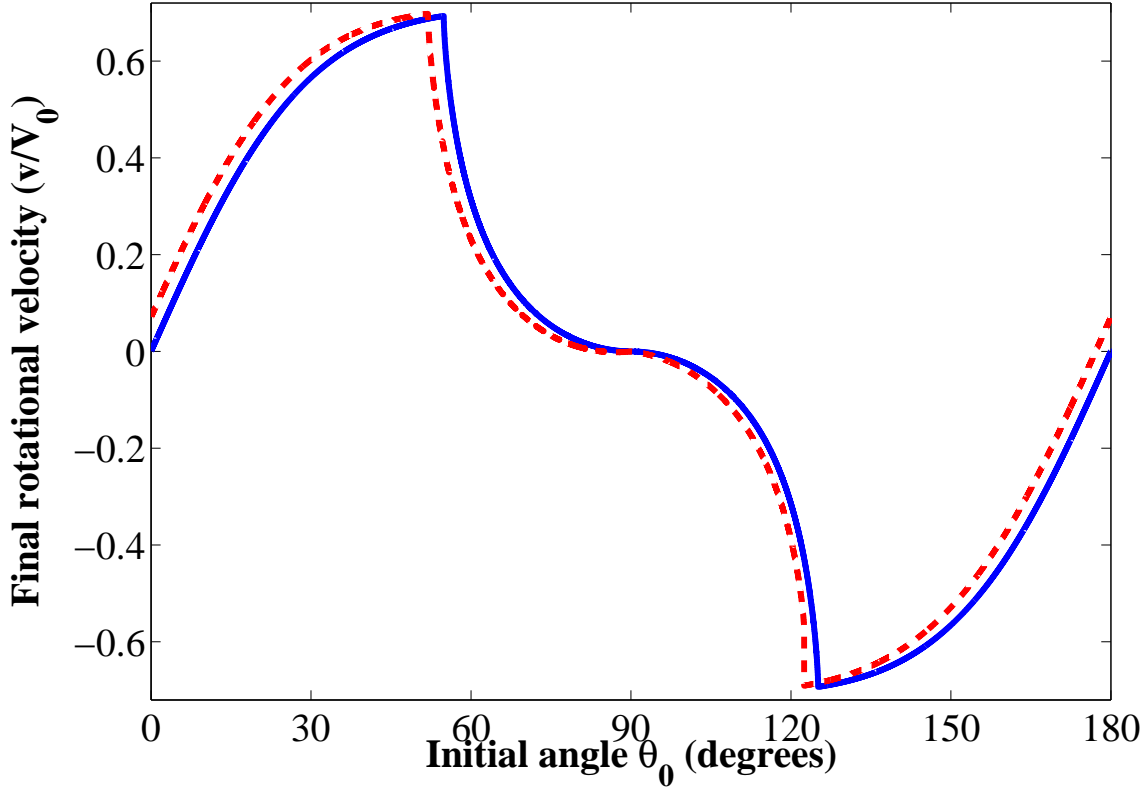}
\caption{(Color online) Rotational velocity of the scattered molecules is plotted as a function of the initial orientation angle of the molecules in the $XZ$-plane. Initially, the molecules are non-rotating, and they impinge on a frozen surface with an incident angle of $\theta_{\text{inc}}=-45^{\circ}$. The solid (blue) line represents scattering from a frictionless surface ($\gamma=0$), while the dashed (red) line is for scattering from a surface with friction ($\gamma=0.1$).}
\label{figure_v_vs_theta}
\end{figure}

\section{Scattering of unidirectional rotating molecules from a surface with friction}
\label{results2}

\begin{figure}
\centering
\includegraphics[width=1\textwidth]{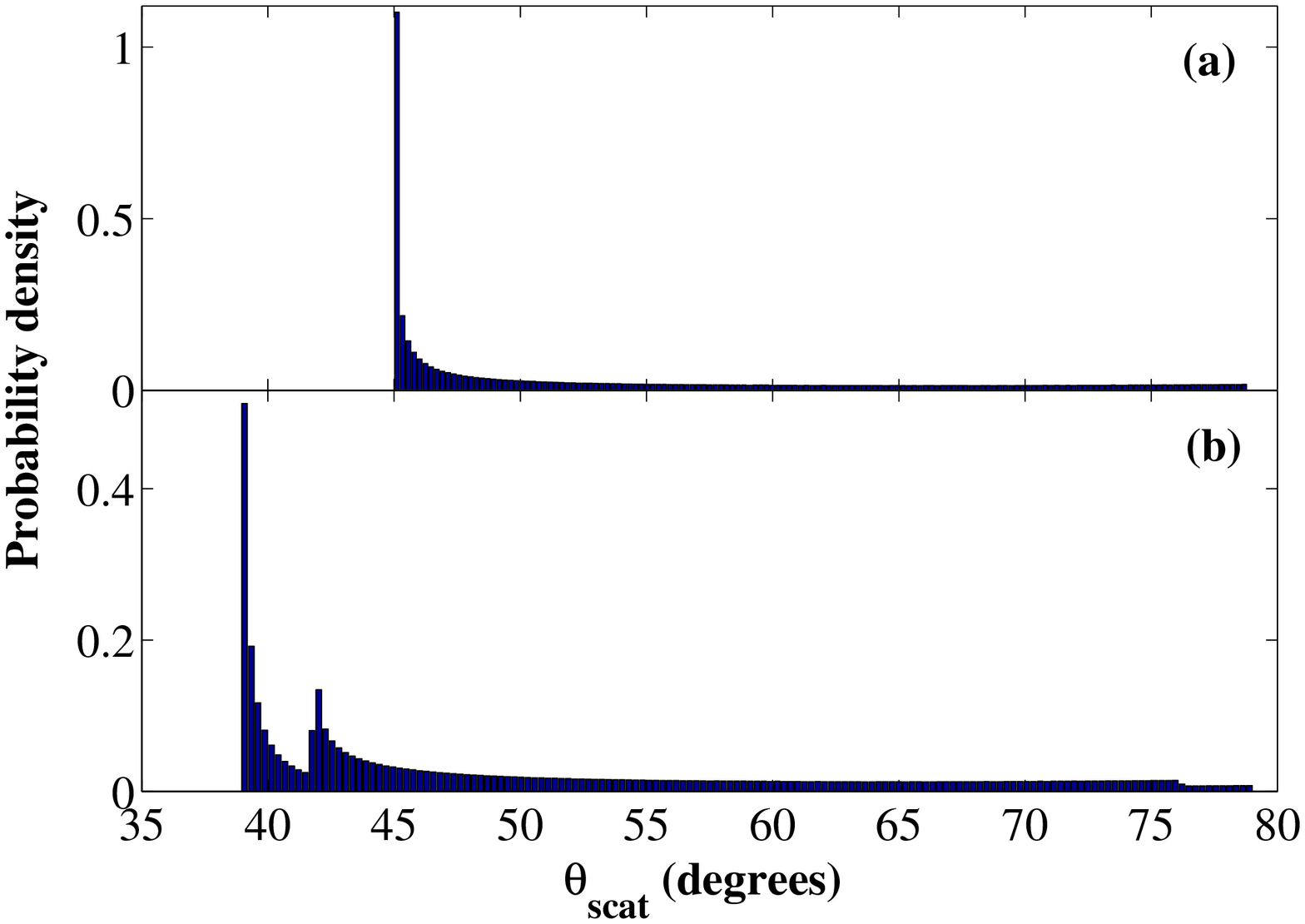}
\caption{The distribution of the scattering angles of $10^5$ molecules scattered from a frozen surface ($T_{\text{surf}}=0$). The molecules are non-rotating and are randomly oriented in the $XZ$-plane before the scattering, and they impinge on the surface with $\theta_{\text{inc}}=-45^{\circ}$. (a) The surface is frictionless with $\gamma=0$, and the scattering is mainly specular, similar to scattering of atoms. (b) The surface is rough with a friction coefficient $\gamma=0.1$, and the scattering is mainly towards angles lower than specular, because the friction of the surface slows down the motion of the molecules in the $X$-direction.}
\label{figure_ang_dist_friction}
\end{figure}
In the current Section, we investigate the influence of the direction of rotation of molecules on their scattering from a surface with friction.
In subsections \ref{results2_nonrot} and \ref{results2_rot_2d} we consider some idealized cases of non-rotating and rotating molecules randomly oriented in the $XZ$-plane, that are scattered from a frozen surface with friction. 
After clarifying the ``clean'' effects seen for two-dimensional, zero temperature surface conditions, we treat the three-dimensional molecular rotation, finite temperature vibrating surface case, with addition of laser pulses, in subsection \ref{results2_rot_3d}.

In all the following subsections, the molecules collide with the surface with friction at an incident angle of $-45^{\circ}$. In subsections \ref{results2_nonrot} and \ref{results2_rot_2d} the values of the velocities are all given in units of $V_0$, such that $V_{X,0}=|V_{Z,0}|=1/\sqrt{2}$.

\subsection{Scattering of non-rotating molecules -- a two-dimensional case}
\label{results2_nonrot}

\begin{figure}
\centering
\includegraphics[width=1\textwidth]{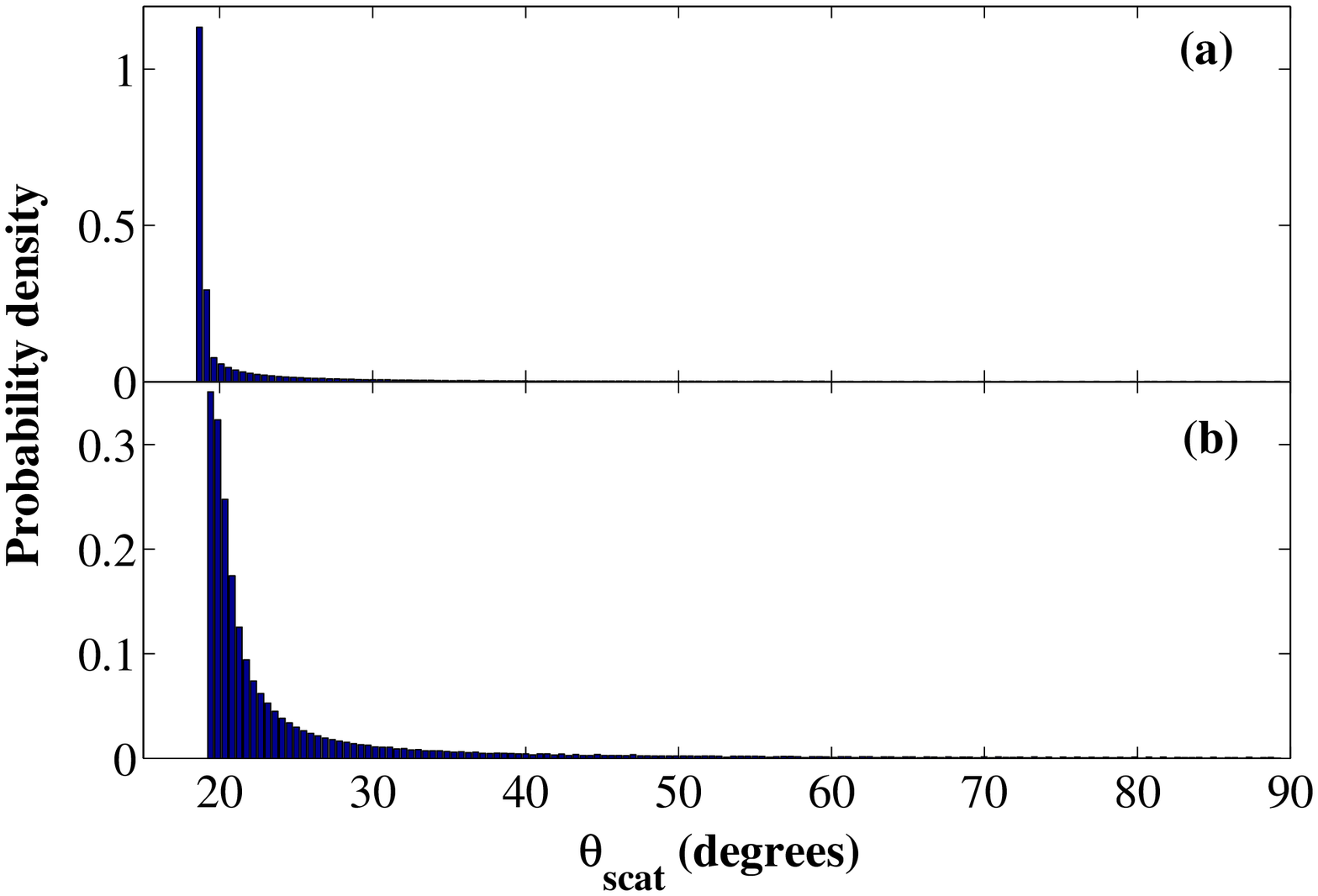}
\caption{The distribution of the scattering angles of $10^5$ molecules scattered from a frozen surface ($T_{\text{surf}}=0$). The molecules are rotating clockwise in the $XZ$-plane before scattering, with $v_0/V_0=2$, and they impinge on the surface with $\theta_{\text{inc}}=-45^{\circ}$. (a) The surface is frictionless ($\gamma=0$), and the molecules scatter mainly to subspecular angles because the energy is transferred from rotational to translational degrees of freedom, during the collision. (b) The surface is rough with a friction coefficient of $\gamma=0.1$. The scattering peak moves towards a slightly higher angle than in (a) because of the frictional force, as explained in the text, and the width of the distribution is significantly higher than in (a).}
\label{figure_ang_dist_friction_rot+}
\end{figure}

In order to check the two-dimensional model of Sec.\ \ref{collision2_2d}, we inspect the final scattering angles and the final rotational velocities for initially non-rotating molecules, oriented in the $XZ$-plane, that scatter from a frozen surface with friction.
In Fig.\ \ref{figure_v_vs_theta} we plot the dependence of the final rotational velocity of the scattered molecule on the initial orientation angle $\theta_0$ in the $XZ$-plane. The solid (blue) curve represents the case of a frictionless surface ($\gamma=0$), while the dashed (red) curve is for a surface with friction ($\gamma=0.1$). Three regions can be identified in the curves of Fig.\ \ref{figure_v_vs_theta}: the first region of positive rotational velocities ($0^{\circ}<\theta_0\lesssim 45^{\circ}$), the second region of negative rotational velocities ($135^{\circ}\lesssim\theta_0<180^{\circ}$), and the third region of the sharp transition between the first two regions, with relatively small rotational velocities ($45^{\circ}\lesssim\theta_0\lesssim135^{\circ}$). The first two regions represent a scattering process by a single collision of the molecule with the surface. It is seen that in the first region the friction speeds up the rotation caused by the collision, while in the second region the friction slows down the rotation. This asymmetry of the friction force can be understood if the strong normal force $\mathbf{F_{\text{n}}}$ and the weak friction force $\mathbf{F_{\text{fr}}}$ are analyzed for the molecule hitting the surface, see Fig.\ \ref{molecule_surface_friction}. The normal force gives some velocity to the molecule, which is symmetric but opposite in sign for molecules inclined by the same angle to the right ($0^{\circ}<\theta_0\lesssim 45^{\circ}$) and to the left ($135^{\circ}\lesssim\theta_0<180^{\circ}$). This can be also seen from the second term on the right of Eq.\ (\ref{v+_pre}). On the other hand, the friction force acts always in the same direction on the molecules falling on the surface from the left (negative incident angles). This way, the friction speeds up the rotation, caused by the normal force, of a molecule inclined to the right, however, it slows down the rotation of a molecule inclined to the left. This can be seen from the third term on the right of Eq.\ (\ref{v+_pre}), which has a positive sign for all the angles $\theta$, for non-rotating molecules. 

The third region of the curves in Fig.\ \ref{figure_v_vs_theta} ($45^{\circ}\lesssim\theta_0\lesssim135^{\circ}$) is a region where the scattering process contains double collisions. Here, the rotational velocity given to the molecule by the first collision is approximately canceled out by the second collision.

In Figure \ref{figure_ang_dist_friction}(a) we plot the scattering angle distribution of non-rotating molecules in the $XZ$-plane scattered from a frictionless surface. The majority of the molecules scatter towards the specular angle around $45^{\circ}$, as expected from Fig.\ \ref{figure_v_vs_theta}. However, from Fig.\ \ref{figure_ang_dist_friction}(b) we see that when friction is added, two peaks appear in the distribution, and they move to smaller scattering angles. The shift to smaller angles can be understood as an effect of the rough surface that ``slows down'' the translational motion of the molecules in the $X$-direction. The peak at $42^{\circ}$ in Fig.\ \ref{figure_ang_dist_friction}(b) corresponds to molecules that collide once with the surface, while oriented at and near $\theta=0^{\circ}$. The higher peak at $39^{\circ}$ corresponds to molecules oriented around $\theta=90^{\circ}$ and colliding twice with the surface. The molecules colliding twice lose more of $V_X$ than the molecules colliding once, and scatter, as a result, to smaller scattering angles.
The location of the peaks for non-rotating molecules can be found according to:
\begin{equation}
\theta_{\text{scat,peak}}=\arctan{\frac{V_{X,f}}{V_{Z,f}}}=\arctan{\frac{(1-n\gamma)V_{X,0}}{|V_{Z,0}|}}~, 
\end{equation}
where $n$ is the number of collisions, that can be equal to $1$ or $2$.
 
\subsection{Scattering of unidirectional rotating molecules -- an idealized two-dimensional case}
\label{results2_rot_2d}

\begin{figure}
\centering
\includegraphics[width=1\textwidth]{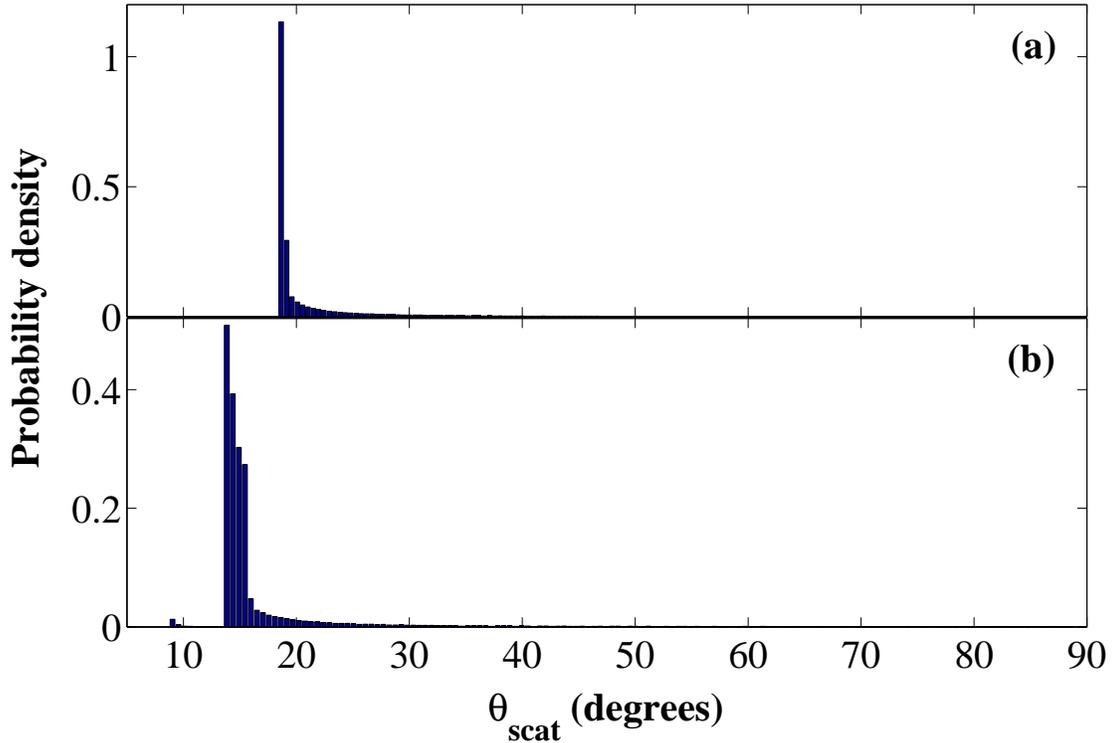}
\caption{The distribution of the scattering angles of $10^5$ molecules scattered from a frozen surface ($T_{\text{surf}}=0$). The molecules are rotating counter-clockwise in the $XZ$-plane before scattering, with $v_0/V_0=-2$. They impinge on the surface with $\theta_{\text{inc}}=-45^{\circ}$. (a) The surface is frictionless ($\gamma=0$). This Figure is the same as Fig.\ \ref{figure_ang_dist_friction_rot+}(a), and is given here for comparison with panel (b). (b) The surface is rough with a friction coefficient of $\gamma=0.1$. The scattering peak moves towards a lower angle than in (a) because of the frictional force, as explained in the text.}
\label{figure_ang_dist_friction_rot-}
\end{figure}

Here we examine the ideal scenario of molecules in the $XZ$-plane, all rotating unidirectionally at a constant angular speed, and colliding with a frozen surface with friction. In Sec.\ \ref{results2_rot_3d}, we extend this scenario to more realistic conditions: molecules set in a three-dimensional rotational motion by femtosecond laser pulses, colliding with a vibrating surface at a finite temperature.

Unidirectionally rotating molecules colliding with a surface with friction can be viewed as a rotating wheel falling on the ground. Inverting the direction of rotation of the wheel should change the scattering process, as explained below.

Imagine a wheel falling on the rough surface from the left, while rotating clockwise. Let us suppose that the rotational velocity of the wheel is high enough, such that the effective velocity of the wheel at the point of contact is directed to the left. In this case, the friction force that is exerted by the surface at impact is directed to the right. This friction force will cause the wheel to scatter towards scattering angles higher than at the absence of friction.
On the other hand, for a wheel that rotates counter-clockwise, the friction force created at the point of contact is directed to the left. This friction force will cause the wheel to scatter towards scattering angles lower than at the absence of friction and closer to the surface normal. 

Now consider molecules initially rotating in the $XZ$-plane with a constant speed of $v_0/V_0=+2$ for the clockwise rotation, and of $v_0/V_0=-2$ for the counter-clockwise rotation, and colliding with the surface with friction.
The effect of the shift in the scattering angles is visible if we compare Figure \ref{figure_ang_dist_friction_rot+}(b) for the clockwise rotation, to Figure \ref{figure_ang_dist_friction_rot-}(b) for the counter-clockwise rotation. 
For comparison, the corresponding scattering angle distributions for rotating molecules scattered from a frictionless surface are shown in Figs.\ \ref{figure_ang_dist_friction_rot+}(a) and \ref{figure_ang_dist_friction_rot-}(a). These two Figures are identical, because the sense of rotation does not change the scattering process for a frictionless surface. A clear deviation of the scattering angle in opposite directions for clockwise and counter-clockwise rotation is seen in Figs.\ \ref{figure_ang_dist_friction_rot+}(b) and \ref{figure_ang_dist_friction_rot-}(b). However, this deviation is lower for the clockwise rotating molecules, than for the counter-clockwise rotating molecules. The reason for this is the difference between the two cases in the local velocity of the colliding atom. The local velocity parallel to the surface of the colliding atom is given by $V_{X,0}+v_{A,x}$. Let us consider a rotating molecule oriented at $0^{\circ}$ colliding with the surface. In the clockwise rotation case the local velocity of this molecule is $V_{X,0}-|v_0|=-1.3$, while in the counter-clockwise rotation case the local velocity is $V_{X,0}+|v_0|=2.7$. The higher (in absolute value) local velocity in the counter-clockwise rotation case leads to a higher friction force, that results in a higher change of the scattering angle. 

\subsection{Scattering of unidirectional rotating molecules -- a realistic three-dimensional case, with laser pulses}
\label{results2_rot_3d}

After demonstrating the difference between the scattering of molecules rotating in an opposite sense for the idealized two-dimensional case, we turn to more realistic conditions, involving the use of laser pulses. 
In this subsection, we model a molecular beam of nitrogen molecules (atomic mass $m=14\,\text{a.u.}$) falling on the surface with friction at an angle of incidence of $-45^{\circ}$ and at a constant speed of $V_0=300\,\text{m/s}$. The molecules in the beam have rotational temperature of $T_{\text{rot}}=5\,\text{K}$, which is typical for molecular beam experiments. The surface is composed of cubes with $M=108\,\text{a.u.}$, corresponding to silver atoms, with oscillation frequency of $\omega_M=10^{13}\,\text{Hz}$, corresponding to the $\text{Ag}(111)$ surface \cite{Doak}. See Sec.\ \ref{vibration} for the definition of the parameters. The surface is at room temperature, $T_{\text{surf}}=300\,\text{K}$, and has a friction coefficient of $\gamma=0.1$.

Before hitting the surface, the molecules are illuminated by two consecutive laser pulses, following the scheme introduced in \cite{Fleischer09}, and experimentally realized in \cite{Kitano09}. See Section \ref{laser} for the rotational velocity change caused by a single laser pulse. The first linearly polarized ultrashort laser pulse causes transient alignment of the molecules. The second pulse, hitting the molecules at the moment of maximal alignment, and polarized at $45^{\circ}$ to the first pulse, causes the molecules to rotate preferentially in the plane defined by the polarization directions of the two pulses. The sense of rotation (clockwise vs.\ counter-clockwise) is defined by the relative orientation of the polarization directions of the first and the second pulse. We want to create a rotation in the $XZ$-plane, and choose the polarization direction of the first pulse to be along the $Z$-axis. This fixes the polarization direction of the second pulse to $\pm 45^{\circ}$ from the $Z$-axis, where the plus sign corresponds to the ignition of the clockwise rotation in the $XZ$-plane, while the minus sign ignites the counter-clockwise rotation. The pulse strength chosen is $P_1=5$ for the first pulse and $P_2=10$ for the second pulse. The value of $P_2$ is higher than that of $P_1$, unlike in \cite{Fleischer09}, in order to create higher rotational velocities.

\begin{figure}
\centering
\includegraphics[width=1\textwidth]{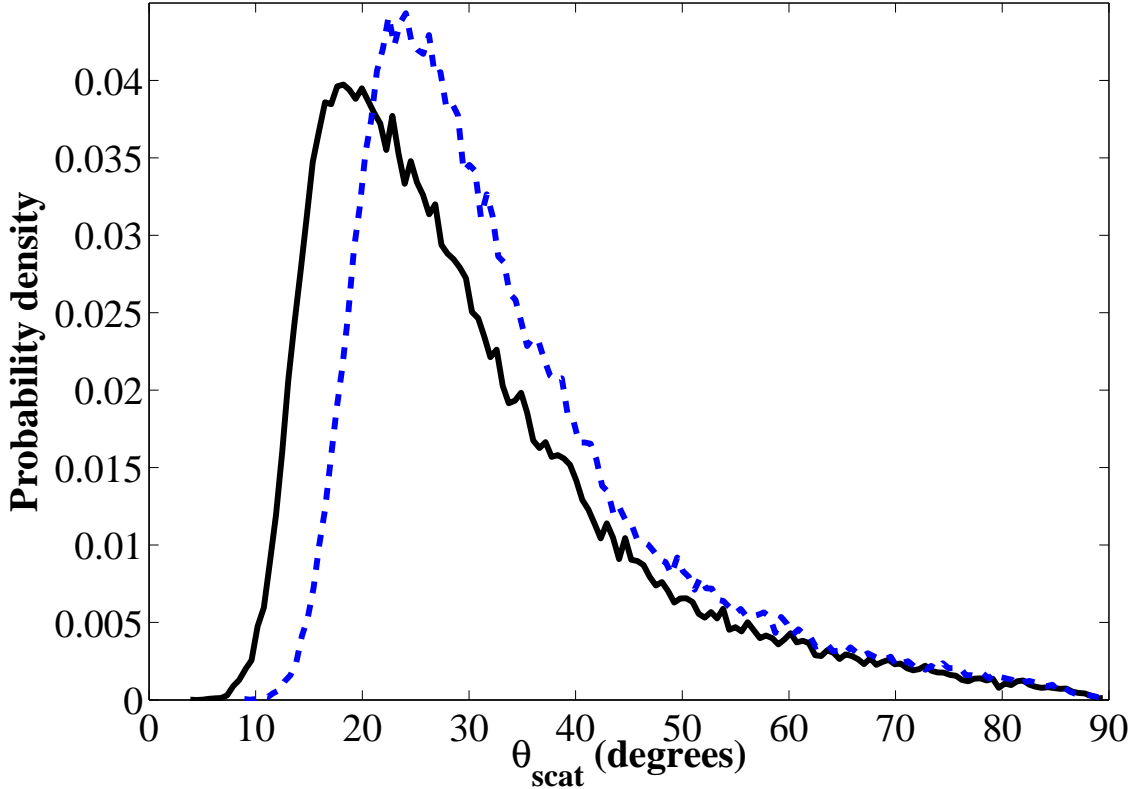}
\caption{(Color online) Two distributions of the scattering angles of molecules set in unidirectional rotation by two laser pulses, as explained in the text. The dashed (blue) line corresponds to clockwise rotating molecules, while the solid (black) line corresponds to counter-clockwise rotating molecules. In this Figure the results of a Monte Carlo simulation of $10^5$ molecules are shown.}
\label{figure_friction_with_laser}
\end{figure}

In Fig.\ \ref{figure_friction_with_laser} we show two distributions of scattering angles, corresponding to the simulations with the parameters above. The dashed line (blue) distribution corresponds to molecules set in a clockwise rotation by the laser pulses, while the solid line (black) distribution corresponds to the counter-clockwise rotating molecules. A distinction of about $5^{\circ}$ can be made between the peaks of the two distributions. The distinction between the oppositely rotating molecules is not as sharp as in the idealized case of Figs.\ \ref{figure_ang_dist_friction_rot+} and \ref{figure_ang_dist_friction_rot-}, and the two distributions in Fig.\ \ref{figure_friction_with_laser} have a significant overlap. However, at the left edge of the distributions a significant contrast exists between the numbers of oppositely rotating scattered molecules. As an example, between the angles of $10^{\circ}$ and $15^{\circ}$ to the surface normal, the probability to find counter-clockwise rotating molecules is about one order of magnitude higher than the probability to find clockwise rotating molecules.

\section{Conclusions}
\label{conclusions}

We introduced two simple classical models, that are both based on the hard cube model, and both treat the diatomic molecule as a dumbbell. The models were used to describe the laser-induced modification of the molecule-surface scattering process. The first model treated the surface as a frictionless surface with a corrugation geometry. The second model treated the surface as flat, with tangential friction forces present.

Using these two models, two potential schemes of laser-controlled surface scattering were introduced and investigated. In the first scheme, the molecules were aligned parallel or perpendicular to the surface by a laser pulse, before hitting the surface. These molecules scattered mainly in two different directions, according to the surface corrugation, and the angle of incidence. The molecules in these two directions were found to be rotating in an opposite sense. This difference in angular momentum between the molecules can, in principle, be detected using resonance-enhanced multiphoton ionization (REMPI) spectroscopy \cite{Sitz88,Kitano09}. In the second scheme, the molecules were prepared in a state with a specific sense of rotation before the scattering from the surface. It was shown that because of the friction with the surface the scattering angles of the rotating molecules depend on their initial direction of rotation. This result, combined with suggestions in \cite{Fleischer09}, can provide new means for separation of molecular mixtures of different species (such as molecular isotopes, or nuclear-spin isomers).
It is our hope, that this work will arouse interest in the femtosecond lasers community, as well as in the molecule-surface scattering field of research. 

\section*{Acknowledgments}

The authors appreciate many years of fruitful discussions with Bretislav Friedrich on various aspects of laser control of molecular alignment.
Y.\ K.\ is thankful to Asaf Azuri and Shauli Daon for stimulating conversations. Financial support of this research by the Israel Science Foundation (Grant No.\ 601/10) is gratefully acknowledged. I.\ A.\ is an incumbent of the Patricia Elman Bildner Professorial Chair. This research is made possible in part by the historic generosity of the Harold Perlman Family.

\end{document}